\documentstyle[preprint,pra,aps]{revtex}
\begin{document}
\draft
\preprint{TAUP 2334-96}
\title{
ACTION AND PASSION AT A DISTANCE\\
An Essay in Honor of Professor Abner Shimony\thanks{
To appear in {\it Quantum Potentiality, Entanglement, and
Passion-at-a-Distance:  Essays for Abner Shimony}, R. S. Cohen,
M. A. Horne and J. Stachel, eds. (Dordrecht, Netherlands:
Kluwer Academic Publishers), in press.}}
\author{Sandu Popescu}
\address{Department of Physics, Boston University, Boston, MA 02215, U.S.A.}
\author{Daniel Rohrlich}
\address{School of Physics and Astronomy,
Tel-Aviv University, Ramat-Aviv 69978 Tel-Aviv, Israel}
\date{\today}
\maketitle
\begin{abstract}
Quantum mechanics permits nonlocality---both nonlocal correlations and
nonlocal equations of motion---while respecting relativistic causality.
Is quantum mechanics the unique theory that reconciles nonlocality and
causality?  We consider two models, going beyond quantum mechanics, of
nonlocality---``superquantum" correlations, and nonlocal ``jamming" of
correlations---and derive new results for the jamming model.  In one
space dimension, jamming allows reversal of the sequence of cause and
effect; in higher dimensions, however, effect never precedes cause.
\end{abstract}
\def\a{\hat a}
\def\ap{\hat a^\prime}
\def\b{\hat b}
\def\bp{\hat b^\prime}
\def\AP{A^\prime}
\def\BP{B^\prime}
\def \ra {\rangle}
\def \la {\langle}
\section{Introduction}
     Why is quantum mechanics what it is?  Many a student has asked this
question.  Some physicists have continued to ask it.  Few have done so
with the passion of Abner Shimony.  ``Why is quantum mechanics what it
is?" we, too, ask ourselves, and of course we haven't got an answer.
But we are working on an answer, and we are honored to dedicate this
work to you, Abner, on your birthday.

     What is the problem?  Quantum mechanics has an axiomatic structure,
exposed by von Neumann, Dirac and others.  The axioms of quantum
mechanics tell us that every state of a system corresponds to a vector
in a complex Hilbert space, every physical observable corresponds to a
linear hermitian operator acting on that Hilbert space, etc.  We see the
problem in comparison with the special theory of relativity.  Special
relativity can be deduced in its entirety from two axioms:  the equivalence
of inertial reference frames, and the constancy of the speed of light.
Both axioms have clear physical meaning.  By contrast, the numerous
axioms of quantum mechanics have no clear physical meaning.  Despite
many attempts, starting with von Neumann, to derive the Hilbert space
structure of quantum mechanics from a ``quantum logic", the new axioms
are hardly more natural than the old.

     Abner Shimony offers hope, and a different approach.  His point
of departure is a remarkable property of quantum mechanics:  nonlocality.
Quantum correlations display a subtle nonlocality.  On the one hand,
as Bell\cite{bell} showed, quantum correlations could not arise in any
theory in which all variables obey relativistic causality\cite{note1}.
On the other hand, quantum correlations themselves obey relativistic
causality---we cannot exploit quantum correlations to transmit signals
at superluminal speeds\cite{c} (or at any speed).  That quantum mechanics
combines nonlocality and causality is wondrous.  Nonlocality and causality
seem {\it prima facie} incompatible.  Einstein's causality contradicts
Newton's action at a distance.  Yet quantum correlations do not permit
action at a distance, and Shimony\cite{s1} has aptly called the nonlocality
manifest in quantum correlations ``passion at a distance".  Shimony has
raised the question whether nonlocality and causality can peacefully
coexist in any other theory besides quantum mechanics\cite{s1,s2}.

     Quantum mechanics also implies nonlocal equations of motion, as Yakir
Aharonov\cite{a1,a2} has pointed out.  In one version of the Aharonov-Bohm
effect\cite{ab}, a solenoid carrying an isolated magnetic flux, inserted
between two slits, shifts the interference pattern of electrons passing
through the slits.  The electrons therefore obey a nonlocal equation of
motion:  they never pass through the flux yet the flux affects their
positions when they reach the screen\cite{note2}.  Aharonov has shown that
the solenoid and the electrons exchange a physical quantity, the {\it
modular momentum}, nonlocally.  In general, modular momentum is measurable
and obeys a nonlocal equation of motion.  But when the flux is constrained
to lie between the slits, its modular momentum is completely uncertain,
and this uncertainty is just sufficient to keep us from seeing a violation
of causality.  Nonlocal equations of motion imply action at a distance,
but quantum mechanics manages to respect relativistic causality.  Still,
nonlocal equations of motion seem so contrary to relativistic causality
that Aharonov\cite{a2} has asked whether quantum mechanics is the {\it
unique} theory combining them.

     The parallel questions raised by Shimony and Aharonov lead us to
consider models for theories, going beyond quantum mechanics, that
reconcile nonlocality and causality.  Is quantum mechanics the only
such theory?  If so, nonlocality and relativistic causality together
imply quantum theory, just as the special theory of relativity can be
deduced in its entirety from two axioms\cite {a2}.  In this paper,
we will discuss model theories\cite{pr,gpr,rp} manifesting nonlocality
while respecting causality.  The first model manifests nonlocality in
the sense of Shimony:  nonlocal correlations.  The second model
manifests nonlocality in the sense of Aharonov:  nonlocal dynamics.
We find that quantum mechanics is {\it not} the only theory that
reconciles nonlocality and relativistic causality.  These models raise
new theoretical and experimental possibilities.  They imply that quantum
mechanics is only one of a class of theories combining nonlocality and
causality; in some sense, it is not even the most nonlocal of such theories.
Our models raise a question:  What is the minimal set of physical
principles---``{\it nonlocality} plus {\it no signalling} plus {\it
something else simple and fundamental}" as Shimony put it\cite{s3}---from
which we may derive quantum mechanics?

\section{Nonlocality I:  nonlocal correlations}
The Clauser, Horne, Shimony, and Holt\cite{chsh} form of Bell's
inequality holds in any classical theory (that is, any theory of local
hidden variables).  It  states that a certain combination of correlations
lies between -2 and 2:
\begin{equation}
-2 \le E(A,B)+E(A,\BP )+E(\AP ,B)-E(\AP ,\BP )\le 2
~~~~.
\label{1}
\end{equation}
Besides 2, two other numbers, $2\sqrt{2}$ and $4$, are important bounds
on the CHSH sum of correlations.  If the four correlations in Eq.\ (\ref{1})
were independent, the absolute value of the sum could be as much as 4. For
quantum correlations, however, the CHSH sum of correlations is bounded
\cite{t} in absolute value by $2\sqrt{2}$. Where does this bound come from?
Rather than asking why quantum correlations violate the CHSH inequality,
we might ask why they do not violate it {\it more}.  Suppose that quantum
nonlocality implies that quantum correlations violate the CHSH inequality
at least sometimes.  We might then guess that relativistic causality is
the reason that quantum correlations do not violate it maximally. Could
relativistic causality restrict the violation to $2\sqrt {2}$ instead of
4?  If so, then nonlocality and causality would together determine the
quantum violation of the CHSH inequality, and we would be closer to a
proof that they determine all of quantum mechanics.  If not, then quantum
mechanics cannot be the unique theory combining nonlocality and causality.
To answer the question, we ask what
restrictions relativistic causality imposes on joint probabilities.
Relativistic causality forbids sending messages faster than light. Thus,
if one observer measures the observable $A$, the probabilities for the
outcomes $A=1$ and $A=-1$ must be independent of whether the other observer
chooses to measure $B$ or $\BP$. However, it can be shown\cite{pr,etc}
that this constraint does not limit the CHSH sum of quantum
correlations to $2\sqrt{2}$. For example, imagine a ``superquantum"
correlation function $E$ for spin measurements along given axes.
Assume $E$ depends only on the relative angle $\theta$ between axes.
For any pair of axes, the outcomes $\vert \uparrow \uparrow \rangle$
and $\vert \downarrow \downarrow \rangle$ are equally likely, and
similarly for $\vert \uparrow \downarrow \rangle$ and $\vert \downarrow
\uparrow \rangle$.  These four probabilities sum to 1, so the probabilities
for $\vert \uparrow \downarrow \rangle$ and $\vert \downarrow \downarrow
\rangle$ sum to $1/2$. In any direction, the probability of $\vert
\uparrow \rangle$ or $\vert \downarrow \rangle$ is $1/2$ irrespective
of a measurement on the other particle.  Measurements on one particle
yield no information about measurements on the other, so relativistic
causality holds.  The correlation function then satisfies $E(\pi -
\theta ) = -E(\theta )$.  Now let $E(\theta )$ have the form

     (i) $E (\theta ) =1$ for $0 \le \theta \le \pi /4$;

     (ii) $E(\theta )$ decreases monotonically and smoothly from 1 to -1
as $\theta$ increases from $\pi /4$ to $3\pi / 4$;

     (iii) $E(\theta) = -1$ for $3\pi /4 \le  \theta \le \pi$.

Consider four measurements along axes defined by unit vectors $\ap$, $\b$,
$\a$, and $\bp$ separated by successive angles of $\pi /4$ and lying in a
plane. If we now apply the CHSH inequality Eq.\ (\ref{1}) to these
directions, we find that the sum of correlations
\begin{equation}
E(\a , \b ) +E(\ap , \b )+E(\a , \bp )-E(\ap , \bp )
=3E(\pi /4 ) - E(3\pi /4) = 4
\end{equation}
violates the CHSH inequality with the maximal value 4.  Thus, a correlation
function could satisfy relativistic causality and still violate the CHSH
inequality with the maximal value 4.

\section{Nonlocality II: nonlocal equations of motion}
     Although quantum mechanics is not the unique theory combining
causality and nonlocal correlations, could it be the unique theory
combining causality and nonlocal equations of motion?  Perhaps the
nonlocality in quantum dynamics has deeper physical signficance.  Here
we consider a model that in a sense combines the two forms of
nonlocality:  nonlocal equations of motion where one of the physical
variables is a nonlocal correlation.  {\it
Jamming}, discussed by Grunhaus, Popescu and Rohrlich\cite{gpr} is such
a model.  The jamming paradigm involves three experimenters.  Two
experimenters, call them Alice and Bob, make measurements on systems that
have locally interacted in the past.  Alice's measurements are spacelike
separated from Bob's.  A third experimenter, Jim (the jammer), presses a
button on a black box.  This event is spacelike separated from Alice's
measurements and from Bob's.  The black box acts at a distance on the
correlations between the two sets of systems.  For the sake of definiteness,
let us assume that the systems are pairs of spin-1/2 particles entangled
in a singlet state, and that the measurements of Alice and Bob yield
violations of the CHSH inequality, in the absence of jamming; but when
there is jamming, their measurements yield classical correlations (no
violations of the CHSH inequality).

     Indeed, Shimony\cite{s1} considered such a paradigm in the context of
the experiment of Aspect, Dalibard, and Roger\cite{adr}.  To probe the
implications of certain hidden-variable theories\cite{hv}, he wrote, ``Suppose
that in the interval after the commutators of that experiment have been
actuated, but before the polarization analysis of the photons has been
completed, a strong burst of laser light is propagated transverse to but
intersecting the paths of the propagating photons....  Because of the
nonlinearity of the fundamental material medium which has been postulated
[in these models], this burst would be expected to generate excitations,
which could conceivably interfere with the nonlocal propagation that is
responsible for polarization correlations."  Thus, Shimony asked whether
certain hidden-variable theories would predict classical correlations
after such a burst.  (Quantum mechanics, of course, does not.)

     Here, our concern is not with hidden-variable theories or with a
mechanism for jamming; rather, we ask whether such a nonlocal equation
of motion (or one, say, allowing the third experimenter nonlocally to
create, rather than jam, nonlocal correlations) could respect causality.
The jamming model\cite{gpr} addresses this question. In general,
jamming would allow Jim to send superluminal signals.  But remarkably,
some forms of jamming would not; Jim could tamper with nonlocal
correlations without violating causality.  Jamming preserves causality
if it satisfies two constraints, the {\it unary} condition and the {\it
binary} condition.  The unary condition states that Jim cannot use
jamming to send a superluminal signal that Alice (or Bob), by examining
her (or his) results alone, could read.  To satisfy this condition,
let us assume that Alice and Bob each measure zero average spin along
any axis, with or without jamming. In order to preserve causality,
jamming must affect correlations only,
not average measured values for one spin component.  The binary condition
states that Jim cannot use jamming to send a signal that Alice {and}
Bob {\it together} could read by comparing their results, if they could
do so in less time than would be required for a light signal to reach
the place where they meet and compare results.  This condition restricts
spacetime configurations for jamming.  Let $a$, $b$ and $j$ denote the
three events generated by Alice, Bob, and Jim, respectively: $a$ denotes
Alice's measurements, $b$ denotes Bob's, and $j$ denotes Jim's pressing of
the button.  To satisfy the binary condition, the overlap of the forward
light cones of $a$ and $b$ must lie entirely {\it within} the forward light
cone of $j$.  The reason is that Alice and Bob can compare their results
only in the overlap of their forward light cones.  If this overlap is
entirely contained in the forward light cone of $j$, then a light signal
from $j$ can reach any point in spacetime where Alice and Bob can compare
their results.  This restriction on jamming configurations also rules
out another violation of the unary condition.  If Jim could obtain the
results of Alice's measurements prior to deciding whether to press the
button, he could send a superluminal signal to Bob by {\it selectively}
jamming\cite{gpr}.

\section{An effect can precede its cause!}
     If jamming satisfies the unary and binary conditions, it preserves
causality.  These conditions restrict but do not preclude jamming.  There
are configurations with spacelike separated $a$, $b$ and $j$ that satisfy
the unary and binary conditions.  We conclude that quantum mechanics is
not the only theory combining nonlocal equations of motion with causality.
In this section we consider another remarkable aspect of jamming, which
concerns the time sequence of the events $a$, $b$ and $j$ defined above.
The unary and binary conditions are manifestly Lorentz invariant, but the
time sequence of the events $a$, $b$ and $j$ is not.  A time sequence $a$,
$j$, $b$ in one Lorentz frame may transform into $b$, $j$, $a$ in another
Lorentz frame.  Furthermore, the jamming model presents us with reversals
of the sequence of {\it cause} and {\it effect}:  while $j$ may precede
both $a$ and $b$ in one Lorentz frame, in another frame both $a$ and $b$
may precede $j$.

     To see how jamming can reverse the sequence of cause and effect, we
specialize to the case of one space dimension.  Since $a$ and $b$
are spacelike separated, there is a Lorentz frame in which they are
simultaneous.  Choosing this frame and the pair $(x,t)$ as coordinates
for space and time, respectively, we assign $a$ to the point (-1,0) and
$b$ to the point (1,0).  What are possible points at which $j$ can cause
jamming?  The  answer is given by the binary condition.  It is particularly
easy to apply the binary condition in 1+1 dimensions, since in 1+1
dimensions the overlap of two light cones is itself a light cone.  The
overlap of the two forward light cones of $a$ and $b$ is the forward
light cone issuing from (0,1), so the jammer, Jim, may act as late as
$\Delta t =1$ {\it after Alice and Bob have completed their measurements}
and still jam their results.  More generally, the binary condition
allows us to place $j$ anywhere in the backward light cone of (0,1)
that is also in the forward light cone of (0,-1), but not on the boundaries
of this region, since we assume that $a$, $b$ and $j$ are mutually
spacelike separated.  (In particular, $j$ cannot be at (0,1) itself.)

     Such reversals may boggle the mind, but they do not lead to any
inconsistency as long as they do not generate self-contradictory causal
loops\cite{dbohm,a3}.  Consistency and causality are intimately related.
We have used the term {\it relativistic causality} for the constraint
that others call {\it no signalling}.  What is causal about this
constraint?  Suppose that an event (a ``cause") could influence another
event (an ``effect") at a spacelike separation.  In one Lorentz frame
the cause precedes the effect, but in some other Lorentz frame the effect
precedes the cause; and if an effect can precede its cause, the effect
could react back on the cause, at a still earlier time, in such a way as
to prevent it.  A self-contradictory causal loop could arise.  A man
could kill his parents before they met.  Relativistic causality prevents
such causal contradictions\cite{dbohm}.  Jamming {\it allows} an event
to precede its cause, but does not allow self-contradictory causal loops.
It is not hard to show\cite{gpr} that if jamming satisfies the unary and
binary conditions, it does not lead to self-contradictory causal loops,
regardless of the number of jammers.  Thus, the reversal of the sequence
of cause and effect in jamming is consistent.  It is, however, sufficiently
remarkable to warrant further comment below, and we also show that the
sequence of cause and effect in jamming depends on the space dimension
in a surprising way.

     The unary and binary conditions restrict the possible jamming
configurations; however, they do not require that jamming be allowed for
all configurations satisfying the two conditions.  Nevertheless, we have
made the natural assumption that jamming is allowed for all such
configurations.  This assumption is manifestly Lorentz invariant.
It allows $a$ and $b$ to both precede $j$.  In a sense, it means that
Jim acts along the backward light cone of $j$; whenever $a$ and $b$
are outside the backward light cone of $j$ and fulfill the unary and
binary conditions, jamming occurs.

\section{An effect can precede its cause??}
     That Jim may act after Alice and Bob have completed their measurements
(in the given Lorentz frame) is what may boggle the mind.  How can Jim
change his own past?  We may also put the question in a different way.
Once Alice and Bob have completed their measurements, there can after all
be no doubt about whether or not their correlations have been jammed;
Alice and Bob cannot compare their results and find out until after Jim
has already acted, but whether or not jamming has taken place is already
an immutable fact.  This fact apparently contradicts the assumption that
Jim is a free agent, i.e. that he can freely choose whether or not to jam.
If Alice and Bob have completed their measurements, Jim is {\it not} a free
agent:  he must push the button, or not push it, in accordance with the
results of Alice and Bob's measurements.

     We may be uncomfortable even if Jim acts before Alice and Bob have
both completed their measurements, because the time sequence of the events
$a$, $b$ and $j$ is not Lorentz invariant; $a$, $j$, $b$ in one Lorentz
frame may transform to $b$, $j$, $a$ in another.  The reversal in the
time sequences does not lead to a contradiction because the effect cannot
be isolated to a single spacetime event:  there is no observable effect
at either $a$ or $b$, only correlations between $a$ and $b$ are changed.
All the same, if we assume that Jim acts on either Alice or Bob---whoever
measures later---we conclude he could not have acted on either of them,
because both come earlier in some Lorentz frame.

     What, then, do we make of cause and effect in the jamming
model?  We offer two points of view on this question.  One point of
view is that we don't have to worry; jamming does not lead to any
causal paradoxes, and that is all that matters.  Of course, experience
teaches that causes precede their effects.  Yet experience also teaches
that causes and effects are locally related. In jamming, causes and
effects are nonlocally related.  So we cannot assume that causes must
precede their effects; it is contrary to the spirit of special relativity
to impose such a demand.  Indeed, it is contrary to the spirit of general
relativity to assign absolute meaning to any sequence of three mutually
spacelike separated events, even when such a sequence has a
Lorentz-invariant meaning in special relativity \cite{a3}.  We only demand
that no sequence of causes and effects close upon itself, for a closed
causal loop---a time-travel paradox---would be self-contradictory. If
an effect can precede its cause {\it and both are spacetime events}, then
a closed causal loop can arise.  But in jamming, the cause is a spacetime
event and the effect involves two spacelike separated events; no closed
causal loop can arise\cite{gpr}.

     This point of view interprets cause and effect in jamming as
Lorentz invariant; observers in all Lorentz frames agree that jamming is
the effect and Jim's action is the cause.  A second point of view asks
whether the jamming model could have any other interpretation.  In
a world with jamming, might observers in different Lorentz frames give
different accounts of jamming?  Could a sequence $a$, $j$, $b$ have a
covariant interpretation, with two observers coming to different
conclusions about which measurements were affected by Jim?  (No experiment
could ever prove one of them wrong and the other right\cite{note3}.)
Likewise, perhaps observers in a Lorentz frame where both $a$ and $b$
precede $j$ would interpret jamming as a form of {\it telesthesia}:
Jim knows
whether the correlations measured by Alice and Bob are nonlocal before
he could have received both sets of results.  We must assume, however,
that observers in such a world would notice that jamming always turns
out to benefit Jim; they would not interpret jamming as mere telesthesia,
so the jamming model could not have this covariant interpretation.

     Finally, we note that a question of interpreting cause and effect
arises in quantum mechanics, as well.  Consider the measurements of Alice
and Bob in the absence of jamming.  Their measured results do not indicate
any relation of cause and effect between Alice and Bob; Alice can do nothing
to affect Bob's results, and vice versa.  According to the conventional
interpretation of quantum mechanics, however, the first measurement on a
pair of particles entangled in a singlet state causes collapse of the
state.  The question whether Alice or Bob caused the collapse of the
singlet state has no Lorentz-invariant answer\cite{gpr,aa}.

\section{Jamming in more than one space dimension}
     After arguing that jamming is consistent even if it allows
reversals of the sequence of cause and effect, we open this section
with a surprise:  such reversals arise only in one space dimension!
In higher dimensions, the binary condition itself eliminates such
configurations; jamming is not possible if both $a$ and $b$ precede
$j$.  To prove this result, we first consider the case of 2+1
dimensions.  We choose coordinates $(x,y,t)$ and, as before, place $a$
and $b$ on the $x$-axis, at (-1,0,0) and (1,0,0), respectively.  Let
$A$, $B$ and $J$ denote the forward light cones of $a$, $b$ and $j$,
respectively.  The surfaces of $A$ and $B$ intersect in a hyperbola
in the $yt$-plane.  To satisfy the binary condition, the intersection
of $A$ and $B$ must lie entirely within $J$.  Suppose that this
condition is fulfilled, and now we move $j$ so that the intersection of
$A$ and $B$ ceases to lie within $J$.  The intersection of $A$ and
$B$ ceases to lie within $J$ when its surface touches the surface of
$J$.  Either a point on the hyperbola, or a point on the surface of
either $A$ of $B$ alone, may touch the surface of $J$.  However, the
surfaces of $A$ and $J$ can touch only along a null line (and likewise
for $B$ and $J$); that is, only if $j$ is not spacelike separated
from either $a$ or $b$, contrary to our assumption.  Therefore the only
new constraint on $j$ is that the hyperbola formed by the intersection
of the surfaces of $A$ and $B$ not touch the surface of $J$.  If we
place $j$ on the $t$-axis, at (0,0,t), the latest time $t$ for which this
condition is fulfilled is when the asymptotes of the hyperbola lie along
the surface of $J$.  They lie along the surface of $J$ when $j$ is the
point (0,0,0).  If $j$ is the point (0,0,0), moving $j$ in either the
$x$- or $y$-direction will cause the hyperbola to intersect the surface
of $J$. We conclude that there is no point $j$, consistent with the binary
condition, with $t$-coordinate greater than 0.  Thus, $j$ cannot succeed
both $a$ and $b$ in any Lorentz frame (although it could succeed one of
them).

     For $n>2$ space dimensions, the proof is similar.  The only
constraint on $j$ arises from the intersection of the surfaces of $A$
and $B$.  At a given time $t$, the surfaces of $A$ and $B$ are
$(n-1)$-spheres of radius $t$ centered, respectively, at $x=-1$ and
$x=1$ on the $x$-axis; these $(n-1)$-spheres intersect in an
$(n-2)$-sphere of radius $(t^2 -1)^{1/2}$ centered at the origin.
This $(n-2)$ sphere lies entirely within an $(n-1)$-sphere of radius
$t$ centered at the origin, and approaches it asymptotically for
$t\rightarrow\infty$.  The $(n-1)$-spheres centered at the origin
are sections of the forward light cone of the origin.  Thus, $j$
cannot occur later than $a$ and $b$.

     We find this result both amusing and odd.  We argued above that
allowing $j$ to succeed both $a$ and $b$ does not entail any inconsistency
and that it is contrary to the spirit of the general theory of relativity
to exclude such configurations for jamming.  Nonetheless, we find that
they are automatically excluded for $n\ge2$.

\section{Conclusions}
     Two related questions of Shimony\cite{s1,s2} and Aharonov\cite{a2}
inspire this essay.  Nonlocality and relativistic causality seem {\it
almost} irreconcilable.  The emphasis is on {\it almost}, because
quantum mechanics does reconcile them, and does so in two different
ways.  But is quantum mechanics the unique theory that does so?  Our
answer is that it is not:  model theories going beyond quantum mechanics,
but respecting causality, allow nonlocality both ways.
We qualify our answer by noting that nonlocality is not completely defined.
Relativistic causality is well defined, but nonlocality in quantum
mechanics includes both nonlocal correlations and nonlocal equations of
motion, and we do not know exactly what kind of nonlocality we are
seeking.  Alternatively, we may ask what additional physical principles
can we impose that will single out quantum mechanics as the unique
theory.  Our ``superquantum" and ``jamming" models open new experimental
and theoretical possibilities.  The superquantum model predicts violations
of the CHSH inequality exceeding quantum violations, consistent with
causality.  The jamming model predicts new effects on quantum correlations
from some mechanism such as the burst of laser light suggested by
Shimony\cite{s1}.  Most interesting are the theoretical possibilities.
They offer hope that we may rediscover quantum mechanics as the unique
theory satisfying a small number of fundamental principles:  causality
plus nonlocality ``plus something else simple and fundamental"\cite{s3}.

\acknowledgments
D. R. acknowledges support from the State of Israel, Ministry of Immigrant
Absorption, Center for Absorption in Science.


\begin{references}
\bibitem{note2}
It is true that the electron interacts {\it locally} with
a vector potential.  However, the vector potential is not a physical
quantity; all physical quantities are gauge invariant.
\end{references}
\end{document}